\documentclass[preprint,amsfonts,showkeys]{revtex4-1}
\usepackage{amssymb}
\usepackage{amsmath}
\usepackage{color}
\usepackage{graphicx}
\usepackage{natbib}
\usepackage{hyperref}
\definecolor{darkgreen}{rgb}{0.1,0.6,0.3}
\definecolor{darkred}{rgb}{0.6,0.3,0.1}
\hypersetup{
    colorlinks=true,       
    linkcolor=blue,          
    citecolor=darkgreen,        
    filecolor=magenta,      
    urlcolor= black           
}

\newcommand{\fig}[1]{Fig.~\ref{fig:#1}}
\newcommand{\Fig}[1]{Figure~\ref{fig:#1}}
\newcommand{\figs}[2]{Figures~\ref{fig:#1},~\ref{fig:#2}}
\newcommand{\eq}[1]{Eq.~\ref{eq:#1}}
\newcommand{\sect}[1]{Section~\ref{sec:#1}}
\newcommand{\app}[1]{App.~\ref{app:#1}}

\begin{document}

\title{Eco-evolutionary dynamics of social dilemmas}

\author{Chaitanya S. Gokhale$^{1,2}$}
\email{gokhale@evolbio.mpg.de}
\author{Christoph Hauert$^3$}

\affiliation{%
$^1$New Zealand Institute for Advanced Study, \\ Massey University, Albany, Private Bag 102904, \\ North Shore Mail Centre, 0745, Auckland, New Zealand}%

\affiliation{%
$^2$Research Group for Theoretical models of Eco-evolutionary Dynamics, Department of Evolutionary Theory, Max Planck Institute for Evolutionary Biology, August Thienemann Str-2, 24306, Pl\"{o}n, Germany}%

\affiliation{%
$^3$Department of Mathematics, University of British Columbia,\\
1984 Mathematics Road, Vancouver BC, Canada V6T 1Z2}%

\begin{abstract}
Social dilemmas are an integral part of social interactions.
Cooperative actions, ranging from secreting extra-cellular products in microbial populations to donating blood in humans, are costly to the actor and hence create an incentive to shirk and avoid the costs.
Nevertheless, cooperation is ubiquitous in nature.
Both costs and benefits often depend non-linearly on the number and types of individuals involved -- as captured by idioms such as `too many cooks spoil the broth' where additional contributions are discounted, or `two heads are better than one' where cooperators synergistically enhance the group benefit. 
Interaction group sizes may depend on the size of the population and hence on ecological processes. 
This results in feedback mechanisms between ecological and evolutionary processes, which jointly affect and determine the evolutionary trajectory. 
Only recently combined eco-evolutionary processes became experimentally tractable in microbial social dilemmas. 
Here we analyse the evolutionary dynamics of non-linear social dilemmas in settings where the population fluctuates in size and the environment changes over time. 
In particular, cooperation is often supported and maintained at high densities through ecological fluctuations. 
Moreover, we find that the combination of the two processes routinely reveals highly complex dynamics, which suggests common occurrence in nature.
\end{abstract}


\maketitle

\section{Introduction}
The theory of evolution is based on Darwinian selection, mutation and drift.
These forces along with neo-Darwinian additions of phenotypic variability, frequency-dependence and, in particular, cooperative interactions within and between species, form the basis for major transitions in evolution \citep{maynard-smith:book:1995a,nowak:Science:2004}. 
Ecological effects such as varying population densities or changing environments are typically assumed to be minimal because they often arise on faster timescales such that only ecological averages matter for evolutionary processes. Consequently, evolutionary and ecological dynamics have been studied independently for long. While this assumption is justified in some situations, it does not apply whenever timescales of ecological and evolutionary dynamics are comparable \citep{day:EL:2007}. 
In such cases, ecological and evolutionary feedback may contribute to the unfolding of the evolutionary process.
Empirically, effects of changes in population size are well documented \citep{dobson:bookchapter:1995,bohannan:AMNAT:1999,hudson:Science:1998,fenner:book:1999,bohannan:EL:2000} and has now lead to a burgeoning field in evolutionary theory, which incorporates ecological variation \citep{may:PRSB:1983,frank:Heredity:1991,heesterbeek:bookchapter:1995,roberts:bookchapter:1995,kirby:PHYTO:1997,gandon:AmNat:2009,salathe:ECL:2005,quigley:PRSB:2012,gokhale:BMCEvolBio:2013,song:BMCEvolBio:2015}.

In particular, the independent study of ecological and evolutionary processes may not be able to capture the complex dynamics that often emerge in the combined system. 
Such potentially rich eco-evolutionary dynamics has been explored theoretically and, more recently, empirically confirmed \citep{post:PTRSB:2009,hanski:PNAS:2011,sanchez:PLoSB:2013}.
Population genetics and adaptive dynamics readily embrace ecological scenarios 
\citep[see e.g.][]{pagie:JTB:1999,aviles:EER:1999,yoshida:Nature:2003,day:book:2005,hauert:JTB:2006a,day:bookchapter:2006,day:EL:2007,lion:JEB:2009,jones:AmNat:2009,gandon:Evolution:2009,wakano:PNAS:2009,cremer:PRE:2011} whereas the traditional focus of evolutionary game theory lies on trait frequencies or constant population sizes \citep{taylor:MB:1978,hofbauer:book:1998,nowak:Nature:2004}. 
Here we propose ways to incorporate intricacies of ecological dynamics along with environmental variation in evolutionary games.

\subsection{Ecological setting}
Evolutionary game dynamics is typically assumed to take place in a population of individuals with fixed types or `strategies', which determine their behaviour in interactions with other members of the population \citep{maynard-smith:Nature:1973,zeeman:JTB:1981}. 
Payoffs determine the success of each strategy. 
Strategies that perform better than the average increase in abundance. 
This is the essence of the replicator equation \citep{hofbauer:book:1998} but neglects that evolutionary changes may alter the population dynamics or vice versa. 
Traditionally the population consists of two strategies whose frequencies are given by $x$ and $y=1-x$. 
In order to incorporate ecological dynamics we assume that $x$ and $y$ are (normalized) densities of the two strategies with $x+y\leq 1$ \citep{hauert:PRSB:2006}. 
Consequently, $z = 1-x-y$ provides a measure for reproductive opportunities, e.g. available space. 
Ecological dynamics is reflected in the change of the population density, $x+y$. The evolutionary dynamics of the strategies is affected by intrinsic changes in population density as well as extrinsic sources such as seasonal fluctuations in the interaction parameters and hence the payoffs. For example, in epidemiology the coevolutionary dynamics of virulence and transmission rate of pathogens depends on ecological parameters of the host population. More specifically, changes in the mortality rate of hosts evokes a direct response in the transmission rate of pathogens while virulence covaries with transmission \citep{day:bookchapter:2006}.
Another approach to implement eco-evolutionary feedback is, for example, to explicitly model spatial structure and the resulting reproductive constraints \citep{lion:JEB:2009,alizon:Evolution:2008,le-gaillard:Evolution:2003,van-baalen:JTB:1998}, which then requires approximations in terms of weak selection or moment closures to derive an analytically tractable framework. In contrast, while our model neglects spatial correlations, it enables a more detailed look at evolutionary consequences arising from intrinsically and extrinsically driven ecological changes.

\subsection{Non-linear social dilemmas}
Social dilemmas occur whenever groups of cooperators perform better than groups of defectors but in mixed groups defectors outcompete cooperators \citep{dawes:ARP:1980}. 
This creates conflicts of interest between the individual and the group. 
In traditional (linear) public goods (PG) interactions cooperators contribute a fixed amount $c>0$ to a common pool, while defectors contribute nothing. 
In a group of size $N$ with $m$ cooperators the payoff for defectors is $P_D(m) = r\, m\, c/N$ where $r>1$ denotes the multiplication factor of cooperative investments and reflects that the public good is a valuable resource. 
Similarly, cooperators receive $P_C(m) = P_D(m) - c = P_D(m-1) + r c/N - c$, where the second equality highlights that cooperators `see' one less cooperator among their co-players and illustrates that the net costs of cooperation are $- r c/N + c$ because a share of the benefits produced by a cooperator returns to itself. 
Therefore, it becomes beneficial to switch to cooperation for large multiplication factors, $r>N$, but defectors nevertheless keep outperforming cooperators in mixed groups. 
The total investment in the PG is based on the number of cooperators in the group but the benefits returned by the common resource may depend non-linearly on the total investments. 
For example, the marginal benefits provided additional cooperators may decrease, which is often termed diminishing returns. Conversely, adding more cooperators could synergistically increase the benefits produced as in economies of scale. 
While well studied in economics \citep{taylor:PoliticalStudies:1982,kollock:ARS:1998,schelling:book:2006} such ideas were touched upon earlier in biology \citep{eshel:AmNat:1988} but only recently have they garnered renewed attention \citep{bach:JTB:2006,hauert:JTB:2006a,wakano:PNAS:2009,pacheco:PRSB:2009,wakano:JTB:2011,archetti:EL:2011,purcell:JTB:2012,pena:JTB:2014,pena:JTB:2015}.

The nonlinearity in PG can be captured by introducing a parameter $\omega$, which rescales the effective value of contributions by cooperators based on the number of cooperators present \citep{hauert:JTB:2006a}. 
Hence, the payoff for defectors, $P_D(m)$, and cooperators, $P_C(m)$, respectively, is given by, 
\begin{subequations}
\label{eq:payoffs}
\begin{align}
P_D(m) &= \frac{r c}N (1 + \omega + \omega^2 + \ldots + \omega^{m-1}) = \frac{r c}N \frac{1-\omega^m}{1-\omega}\\
P_C(m) &= P_D(m) - c = \frac{r c}N \omega (1 + \omega + \ldots + \omega^{m-2}) + \frac{r c}N - c,
\end{align}
\end{subequations}
such that the benefits provided by each additional cooperator are either discounted, $\omega<1$, or synergistically enhanced, $\omega>1$. The classic, linear PG is recovered for $\omega=1$. 
This parametrization provides a general framework for the study of cooperation and recovers all traditional scenarios of social dilemmas as special cases \citep{nowak:Science:2004,hauert:JTB:2006a}.

\section{Eco-evolutionary dynamics}
The overall population density, $x+y$, can grow or shrink from $0$ (extinction) to an absolute maximum of $1$ (normalization). 
The average payoffs of cooperators and defectors, $f_C$ and $f_D$, determine their respective birth rates but individuals can successfully reproduce if reproductive opportunities, $z>0$, are available. 
All individuals are assumed to die at equal and constant rate, $d$. 
Formally, changes in frequencies of cooperators and defectors over time are governed by the following extension of the replicator dynamics \citep{hauert:PRSB:2006},
\begin{subequations}
\label{eq:xyzeqs}
\begin{align}
\dot{x} &= x (z f_C -d) \\
\dot{y} &= y (z f_D -d)\\
\dot{z} &= -\dot{x} - \dot{y}\ = (x+y) d-z (x f_C + y f_D).
\end{align}
\end{subequations}
The average payoffs are calculated following \eq{payoffs}, where the interaction group size depends on the population density (see \app{A}). 
This extends the eco-evolutionary dynamics for the linear PG \citep{hauert:PRSB:2006} to account for discounted or synergistically enhanced accumulation of benefits \citep{hauert:JTB:2006a}. 
The difference in the average fitness between defectors and cooperators, $F(x,z)= f_D -f_C$ is now given by
\begin{align}
\label{eq:fz}
F(x,z) = 1 + (r-1) z^{N-1} - \frac{r}{N} \frac{(1-x (1-\omega))^{N}-z^N}{1-z-x (1-\omega)}
\end{align}
and provides a gradient of selection. 
Note that in the special case of the linear PG, $\omega=1$, \eq{fz} reduces to a function of $z$ alone.

\subsection{Intrinsic fluctuations}
\label{sec:internal}

Homogenous defector populations go extinct but pure cooperator populations can persist and withstand larger death rates $d$ under synergy than discounting (see \app{A1}, \Fig{nodefroots}).
In order to analyze the dynamics in heterogenous populations it is useful to rewrite \eq{xyzeqs} in terms of $z$ and the fraction of cooperators, $f=x/(1-z)$:
\begin{subequations}
\label{eq:fzeqs}
\begin{align}
\dot{f} &= \frac{\dot{x} y - \dot{y} x}{(1-z)^2} = -z f (1-f) F(f,z) \\
\dot{z} &= - (1-z) (f z (r-1) (1-z^{N-1})-d).
\end{align}
\end{subequations}
This change of variable introduces a convenient separation in terms evolutionary and ecological dynamics: evolutionary changes affect strategy abundances and are captured by $f$, whereas ecological dynamics are reflected in changes of the (normalized) population density, $x+y=1-z$. 

\begin{figure}
\begin{center}
\includegraphics[width=0.95\columnwidth]{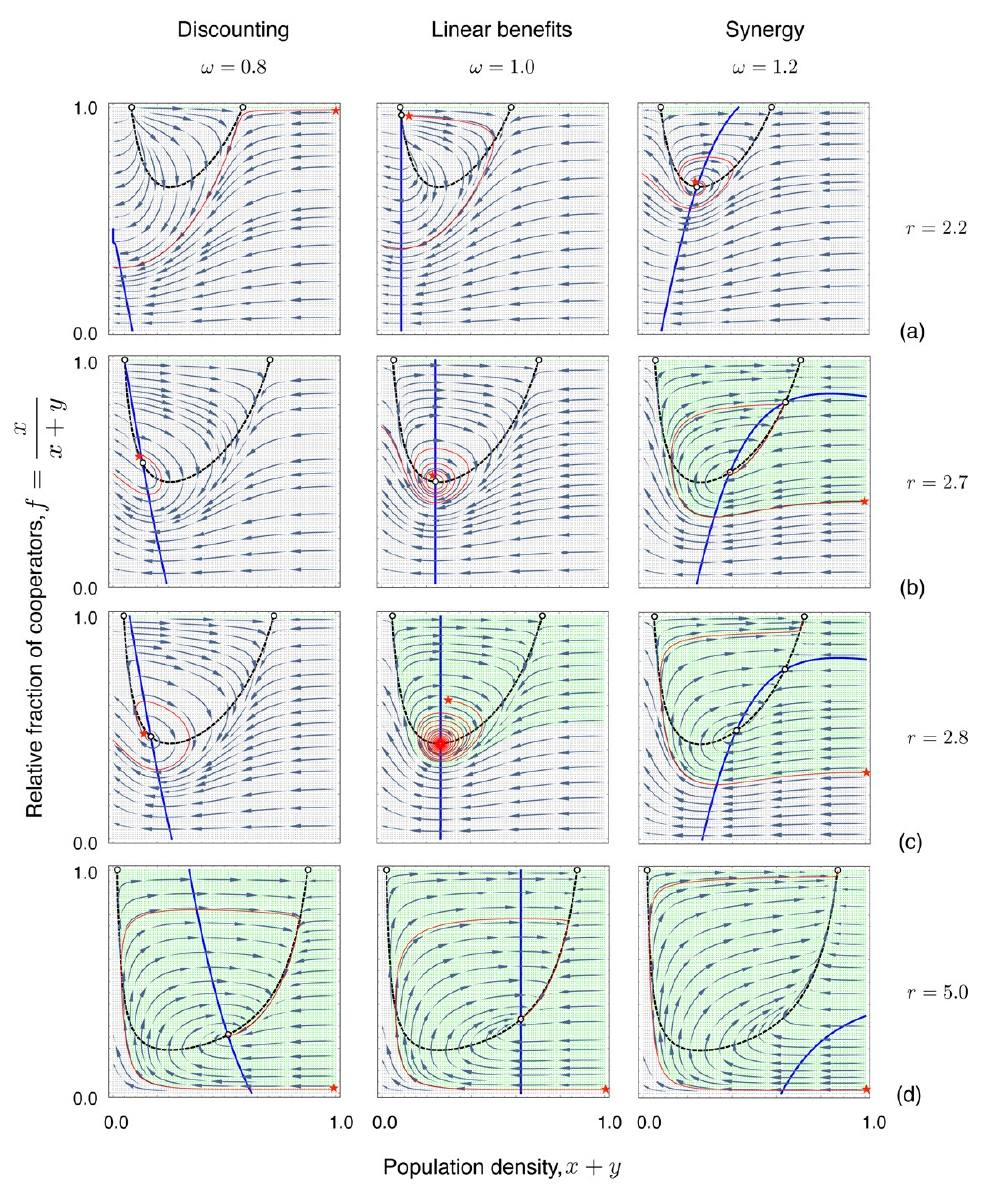}
\caption{
\label{fig:allplots}
\scriptsize{
\textbf{Eco-evolutionary dynamics of public goods interactions with synergy/discounting in the $(1-z,f)$-phase plane.}
Along the null-clines of \eq{fzeqs} the population density ($\dot z=0$, dashed line) or the population composition ($\dot f=0$, solid line) does not change. For $\omega=1$ the null-cline $\dot f=0$ is a vertical line and hence admits at most one interior equilibrium denoted by $\bf{Q}$. 
The non-linearity introduced by synergy and discounting can admit a second interior equilibrium $\bf{P}$.
We set the parameters to $N = 8$, $d=0.5$ and illustrate the dynamics under discounting ($\omega<1$, left column), linear public goods ($\omega=1$, middle column), and synergy ($\omega>1$, right column) for increasing $r$ (top to bottom): (a) $2.2$, (b) $2.7$, (c) $2.8$ and (d) $5.0$.
The stability of the fixed points is discussed in \app{B}.
Example trajectories are shown in red starting at the starred initial configuration.
Compared to the linear public goods (center column), synergistic benefits (right column) admit stable co-existence at lower $r$ whereas higher $r$ are required for discounting (left column).
}
}
\end{center}
\end{figure}

Eco-evolutionary trajectories are visualized in the phase plane $(1-z,f)\in[0,1]^2$. 
In contrast to \cite{hauert:JTB:2002a,hauert:PRSB:2006,hauert:TPB:2008} the interior of the phase plane can support more than one fixed point, because of the non-linearity introduced by synergy/discounting, see \fig{allplots}.
In addition to the equilibria along the boundary, interior equilibria of \eq{fzeqs} are determined by the intersections of the two null-clines given by $f = d/[z (r-1) (1-z^{N-1})]$ ($\dot z=0$) and solutions of $F(f,z) =0$ ($\dot f=0$).
Unfortunately an analytical stability analysis of the interior fixed point(s) is inaccessible but a numerical analysis suggests that a single fixed point, $\bf{Q}$, can exhibit various stability properties depending on $r$, whereas the second fixed point, $\bf{P}$, whenever present, is always a saddle (see \app{B}).

\begin{figure}
\begin{center}
\includegraphics[width=\columnwidth]{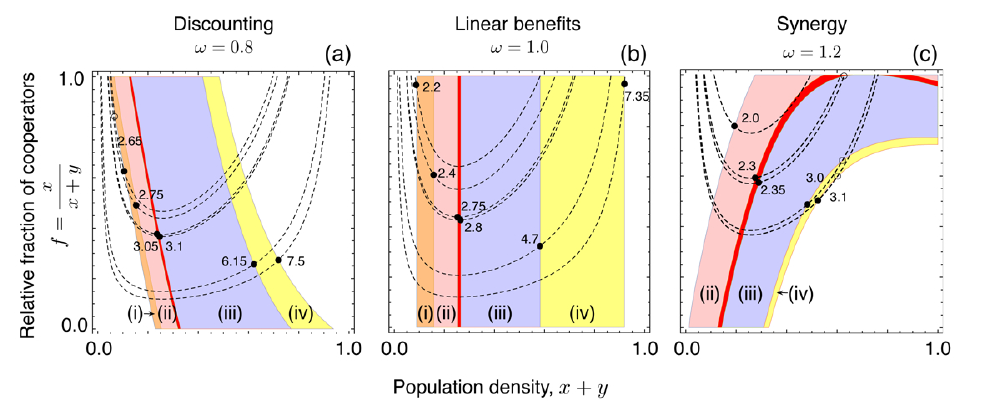}
\caption{
\label{fig:phaseportr}
\scriptsize{
\textbf{Internal fixed points for varying $r$.}
The $z$-null-cline (dashed lines), $f = d/[z (r-1) (1-z^{N-1})]$ is independent of synergy/discounting, $\omega$. The $f$-null-cline (solid lines), is given by the solution of $F(f,z) =0$. The null-clines are shown for specific $r$ (numerically explored range $r\in[2,7.95]$ in increments of $\Delta r=0.05$), where the stability of $\bf{Q}$ (solid circles) changes and separates dynamical regimes: (i) unstable node (orange) (ii) unstable focus (light red) (iii) stable focus (light blue) (iv) stable node (yellow).
Between (ii) and (iii) (red) Andronov-Hopf and other complex bifurcations are possible \citep{hauert:PRSB:2006,hauert:TPB:2008}.
Only under discounting or synergy another fixed point ($\bf{P}$, open circle) may appear. 
\textbf{(a)} Discounting: both internal fixed points appear for $r=2.65$ where $\bf{P}$ is a saddle and $\bf{Q}$ an unstable node. While $\bf{P}$ exits the phase space immediately, $\bf{Q}$ is still present for $r=7.5$.
\textbf{(b)} Linear public goods: only a single internal fixed point ($\bf{Q}$) can exist and for its detailed stability analysis, see \cite{hauert:TPB:2008}.
\textbf{(c)} Synergy: $\bf{Q}$ is already present at $r = 2$ as an unstable focus, $\bf{P}$ appears at $r = 2.35$ as a saddle while $\bf{Q}$ is a stable focus. 
Both fixed points persist until they annihilate each other close to $r = 3.1$. As $\omega$ increases, smaller $r$ are sufficient to render $\bf Q$ stable and prevent extinction. 
}
}
\end{center}
\end{figure}

\begin{figure}
\begin{center}
\includegraphics[width=\columnwidth]{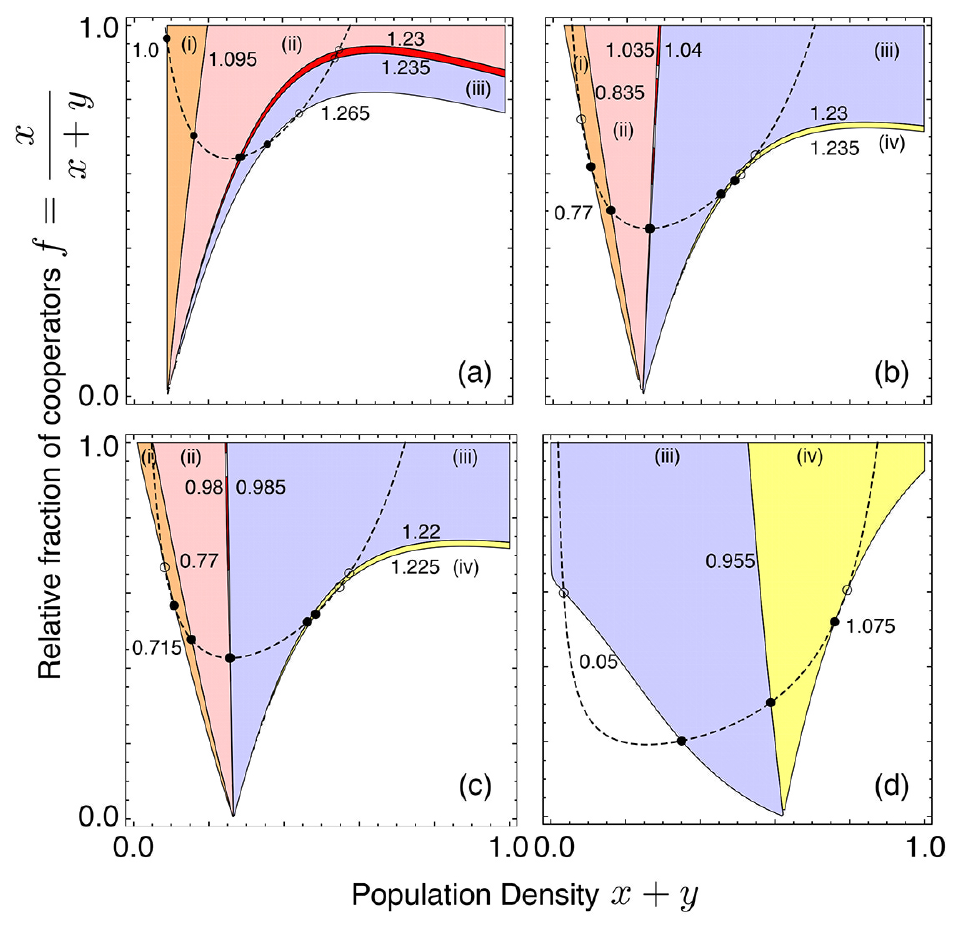}
\caption{
\label{fig:sols}
\scriptsize{
\textbf{Internal fixed points for varying $\omega$.}
The $z$-null-cline ($f = d/[z (r-1) (1-z^{N-1})]$, dashed line), and various $f$-null-cline, ($F(f,z) =0$, solid lines), are shown under synergy/discounting for $\omega\in[0.05,2.0]$ and increments of $\Delta\omega=0.005$. The stability of $\bf Q$ again delineates different dynamical regimes. Note that depending on $r$ not all four regimes may occur as is the case in panels (a) and (d). 
Parameters are $N=8$, $d=0.5$ and $r$ as 
(a) $2.2$, (b) $2.7$, (c) $2.8$ and (d) $5.0$.
(i) at small $\omega$ the fixed point(s) appear (except in (d) where already for $\omega = 0.05$ both fixed points exist). 
Of the two fixed points, one is always a saddle, $\bf{P}$ (open circles), whereas the other, $\bf{Q}$ (solid circles), typically enters the phase plane as an unstable node (orange).
(ii) $\bf{Q}$ becomes an unstable focus (light red). 
Between regions (ii) and (iii) (red) complex bifurcations are possible. 
(iii) $\bf Q$ is a stable focus (light blue).
(iv) $\bf Q$ turns into a stable node (yellow).
For still larger $\omega$ the two interior fixed points collide and annihilate each other. 
}
}
\end{center}
\end{figure}

\Fig{phaseportr} depicts the null-clines for increasing $r$ and fixed $\omega$. The stability of $\bf Q$ defines four dynamical regimes: for small $r$ $\bf Q$ does not exist but when increasing $r$ it (i) appears as an unstable node, (ii) turns into an unstable focus, (iii) then becomes a stable focus and, finally, (iv) a stable node before $\bf Q$ disappears again at high $r$.
Similarly, changing $\omega$ with fixed $r$, triggers a series of bifurcations because the $f$-null-cline depends on $\omega$ whereas the $z$-null-cline does not, see \Fig{sols}.
Note that all $f$-null-clines run through the point defined by $F(0,z)=0$ at the lower boundary of the phase plane. 

A detailed description of the changes in the stability of the fixed point $\bf{Q}$ and hence the eventual dynamics is given in \app{B}.
In \figs{phaseportr}{sols} we describe the dynamics for some fixed values of $r$ and $\omega$, i.e. the rate of return of the public good and the synergy/discounting factor.
However what would happen if the actual values of these parameters changed in a continous fashion over time?

\section{Environmental fluctuations}
\label{sec:external}
A constant feature of evolutionary as well as ecological processes is their dynamic nature. 
However, most evolutionary models assume a deterministic and usually constant environment in which populations evolve -- either deterministically or stochastically \citep{taylor:MB:1978,hofbauer:book:1998,nowak:Nature:2004,moran:book:1962}. 
Considering variable environments is a natural way of incorporating changing ecological conditions. 
Stochastic or periodic fluctuations in the environment may alter evolutionary trajectories as has been shown experimentally \citep{beaumont:Nature:2009}. 
Variable environments have been considered for a variety of interesting phenomena from bet hedging strategies and Red Queen dynamics to the evolution of sex \citep{salathe:TREE:2008,wolinska:TREE:2009}. 
When considering stochastic dynamics, the fixation probability of a trait is a crucial determinant of evolutionary change.
In classical population genetics, it is possible to determine the fixation probability under demographic fluctuations \citep{ewens:book:1979,kimura:PNAS:1974,otto:Genetics:1997}, temporally variable selection strength \citep{jensen:GR:1973,karlin:TPB:1974,uecker:Genetics:2011,carja:TPB:2013} as well as both \citep{waxman:Genetics:2011}, provided that the fitness of traits is frequency independent.
In contrast, in evolutionary games and the evolution of cooperation, in particular, fitnesses are intrinsically frequency dependent and a theoretical understanding of the effects of demographic fluctuations and/or temporally fluctuating game parameters is nascent and has received sporadic attention \citep{uyenoyama:TPB:1979,van-baalen:JTB:1998,hauert:PRSB:2006,alizon:Evolution:2008,wakano:PNAS:2009,lion:JEB:2009,lion:Evolution:2010,lehmann:PTRSB:2010,huang:NatCom:2012}.

Ecological variation can result from feedback between reproductive rates and population densities -- an intrinsic source of variation -- or as a response to extrinsic changes of the environment, which can be implemented by altering the interactions or by varying game parameters.

\subsection{Variation of interaction types}
\begin{figure}
\begin{center}
\includegraphics[width=0.9\columnwidth]{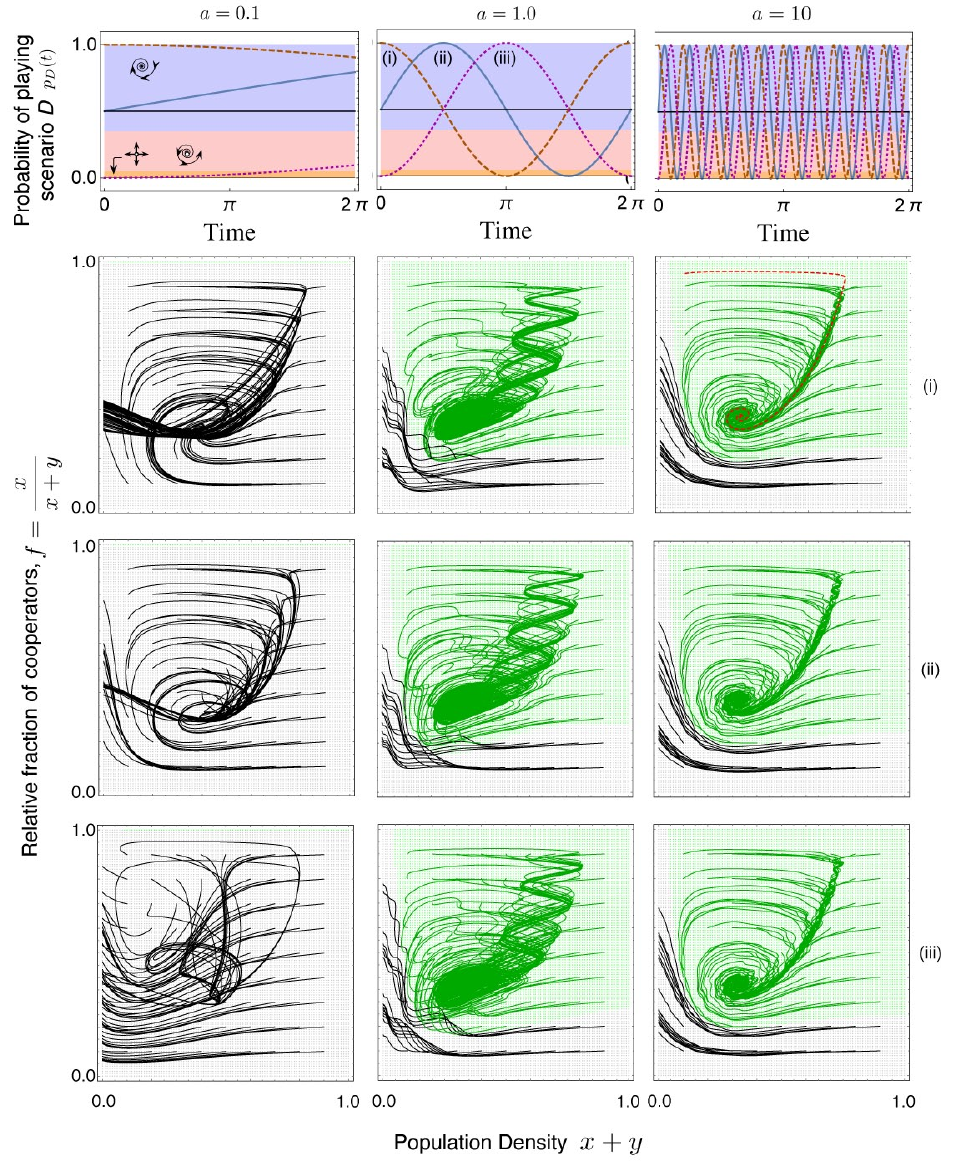}
\caption{
\label{fig:varenv01}
\scriptsize{
\textbf{Eco-evolutionary dynamics under environmental fluctuations: oscillations between interaction types, $\mathbf{p_D(t)}$.}
In scenario $D$ benefits are discounted, $\omega_D=0.9$, with a rate of return of $r_D = 4.2$, as compared to scenario $S$ where benefits are synergistically enhanced, $\omega_S=1.1$, but at a reduced rate of return, $r_S = 2.1$. 
The probability for each type of interaction oscillates over time according to $p_D(t)=( \sin (a t + \delta) + 1)/2$ (top row) with $a=\{0.1,1,10\}$ (columns). The fixed point $\bf Q$ is stable for larger $p_D(t)$ (blue) and unstable for smaller ones (red, orange).
The dynamics for the three phases $\delta = \{\pi/2,0,-\pi/2\}$ labelled (i), (ii) and (iii) are shown in the bottom three rows.
Trajectories are obtained by numerically integrating \eq{fzeqs} with a grain of $0.1$ and those leading to extinction are coloured black while those surviving are green. The background colours are obtained by integrating from initial conditions with a finer grain of $0.01$.
When compared to the average interaction $(D+S)/2$, i.e. when $p_D(t)=0.5$ (\fig{avgdyn}), the panel for $a=1$ and $a=10$ are in good qualitative agreement (sample trajectory (red, dashed) plotted in $a=10$ (i)).
For $a=0.1$ the trajectories follow different paths but all lead eventually to extinction. 
Even when beginning with coexistence (i), this is only transient as $p_D(t)$ eventually renders $\bf{Q}$ unstable and leads the trajectories to extinction from which there is no recovery.
Parameters: $N=8$ and $d=0.5$. 
}
}
\end{center}
\end{figure}

In order to mimic seasonal variation, for example, consider two types of PG interactions, discounting ($D$) and synergy ($S$), both with $N=8$ and $d=0.5$: In $D$ benefits are discounted by $\omega_D = 0.9$ but the multiplication factor is high, $r_D=4.2$, whereas in $S$ benefits are synergistically enhanced by $\omega_S = 1.1$ but for a lower multiplication factor, $r_S=2.1$.
This combination of parameter values ensures that groups of cooperators have the same fitness in $D$ and $S$.
With probability $p_D(t)=(\sin(a t + \delta) + 1)/2$ individuals engage in $D$ and with $p_S(t)=1-p_D(t)$ in $S$, i.e the probability to engage in one or the other type of interaction changes over time reflecting changes in resource abundance or relating to seasonal tasks. 
The parameter $a$ indicates the relation between the timescales of environmental fluctuations and the eco-evolutionary dynamics while $\delta$ tunes the phase of the oscillating wave. 
For $a>1$ environmental fluctuations are faster than the eco-evolutionary dynamics, slower for $a<1$, and for $a=1$ the two timescales are the same.
The dynamical equations, \eq{fzeqs}, thus become:
\begin{subequations}
\label{eq:alteredfz}
\begin{align}
\dot{f} =&\ -z f (1-f) \big[p_D(t) F(f,z,r_D,\omega_D) + p_S(t) F(f,z,r_S,\omega_S)\big]\\
\dot{z} =&\ - (1-z) \big[ z f \left((p_D(t) r_D + p_S(t) r_S) -1\right) (1-z^{N-1})-d\big].
\end{align}
\end{subequations}
The gradient of selection, $F(f,z)$, is split into $F(f,z,r_S,\omega_S)$ and $F(f,z,r_D,\omega_D)$ for the two types of interactions. 
Based on numerical integration, the trajectories of periodic fluctuations between $D$ and $S$ reveal qualitatively similar dynamical properties as the average interaction, $(D+S)/2$, provided that environmental fluctuations are sufficiently fast ($a>1$) (compare \fig{varenv01} with \app{C} and \fig{avgdyn}).

As $p_D(t)$ oscillates the location and even stability of the fixed point $\bf Q$ changes periodically, see \fig{varenv01} for $\omega_D=0.9$, $\omega_S=1.1$, $r_D=4.2$, $r_S=2.1$, $N=8$, and $d=0.5$.
In the pure $D$ scenario, $p_D(t)=1$, $\bf{Q}$ is stable, while for pure $S$, $p_D(t)=0$, $\bf{Q}$ changes position and is unstable (manipulate both at \citep{gokhale:WD:2014}). 
In order to illustrate the detailed dynamics, we consider the stability of $\bf Q$ as a function of $p_D$, which is indicated by the colours and cartoons in the top row of \fig{varenv01}.
For small $p_D$ ($S$ scenario), the fixed point $\bf Q$ is unstable (going from an unstable node, orange to an unstable focus, pink) but becomes stable for $p_D\geq0.35$ (as a stable focus, blue).
For slow oscillations ($a\ll1$) the population invariably goes extinct because $\bf Q$ remains unstable for extended periods. 
Note that in this case the initial condition might matter, i.e. whether the $S$ or $D$ scenario applies first. While this indeed affects the trajectories it does not alter the eventual outcome, see \fig{varenv01}.
In contrast, for comparable timescales ($a\approx1$) or fast oscillations ($a\gg1$), $\bf Q$ is moving fast and changes stability frequently with the net effect that trajectories are attracted towards $\bf Q$ and the population manages to escape extinction.
More specifically, for $a\approx1$ or larger, the coexistence region essentially reflects the averaged case but as $a$ decreases, more and more initial conditions lead to extinction. 
Conversely, for larger $a$, the sizes of the two basins seem no longer affected for large $a$ (comparing $a=1$ and $a=10$), which suggests an upper limit for the basin leading to extinction. 

\subsection{Variation of the rate of return}
Changes in the richness of biological environments, or the economic situation of governing bodies in social settings, can be captured by varying the multiplication factor $r$.
For the traditional, linear PG game ($\omega = 1$) we consider $r(t)=3 \sin (at+\delta) + 4.5$, which ensures $1<r(t)<N$ for $N=8$. As before, $a$ relates the timescales of the eco-evolutionary dynamics and environmental fluctuations and $\delta$ to the phase.

\begin{figure}
\begin{center}
\includegraphics[width=\columnwidth]{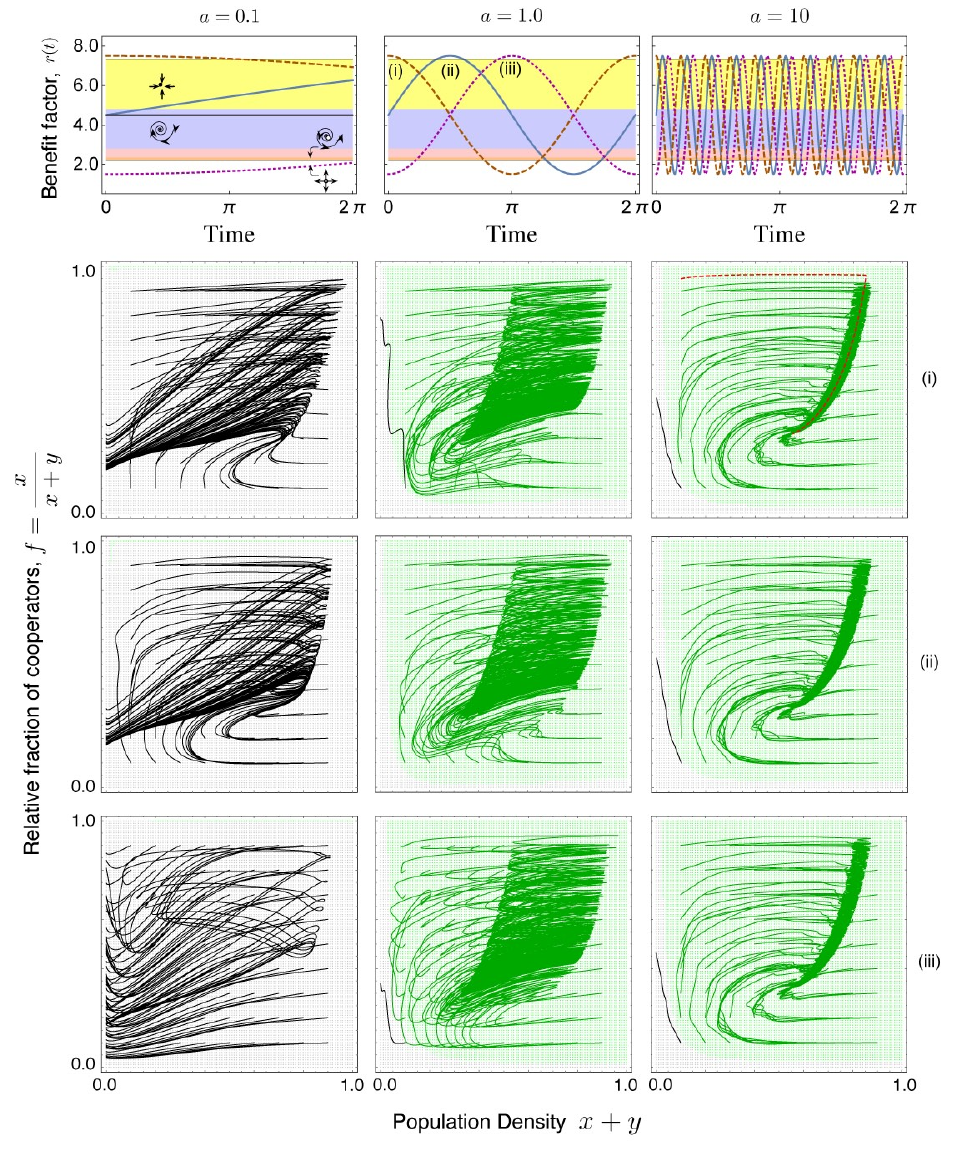}
\caption{
\label{fig:varenv02}
\scriptsize{
\textbf{Eco-evolutionary dynamics under environmental fluctuations: variable rates of return, $\mathbf{r(t)}$.}
The dynamics are depicted for oscillating rates of return, $r(t)=3 \sin (at+\delta) + 4.5$ (top row), three timescales $a=\{0.1,1.0,10\}$ (columns), and three phases $\delta=\{\pi/2,0,-\pi/2\}$ (last three rows). The fixed point $\bf Q$ is stable if $r(t)$ lies in the yellow or blue regions, stable for red or orange and absent in white regions.
The trajectories for comparable and fast ecological timescales are again in good agreement with the dynamics based on the average return, $\bar r=4.5$, which exhibits coexistence (\fig{avgdyn}).
A sample trajectory (red, dashed) for $\bar r=4.5$ is plotted for $a=10$ (i).
For comparable timescales ($a=1$) trajectories oscillate in response to the changing location and stability of $\bf Q$, whereas for fast oscillations environmental changes occur faster than the population can react, which results in an averaging effect.
For slower oscillations extinction is inevitable but the initial phase of $r(t)$ determines the trajectory and thus the time to extinction.
In particular, starting with a high rate of return, $(i)$, the population persists for a longer time.
Parameters: $N=8$ and $d=0.5$. 
}
}
\end{center}
\end{figure}

Again, for comparable timescales ($a\approx1$), or fast oscillations ($a\gg1$), the qualitative dynamics is well captured by the average multiplication factor, $\bar r=4.5$, which also extends to non-linear PG's ($\omega\neq 1$).
Observing the dynamics at fixed, incremental values of $r$ from $1$ to $8$ in discrete steps of $\Delta r = 0.05$, we find that for small $r$ the fixed point $\bf Q$ is missing and extinction is inevitable. 
$\bf{Q}$ only appears at $r = 2.2$ but is still unstable. 
Only for $r = 2.8 $, $\bf{Q}$ turns into a stable fixed point and renders co-existence possible up to $r = 7.35$ at which point $\bf Q$ disappears again and homogenous cooperation becomes a possible stable outcome for most of the initial conditions.
However, the trajectories for the oscillating $r(t)$ are strikingly different, see \fig{varenv02} -- extinction for slow oscillations (small $a$), oscillating trajectories for $a\approx1$ and co-existence for fast oscillations (large $a$).

\subsection{Variation of synergy/discounting}
\label{sec:flucomega}
In order to mimic marginal benefits of joining PG interactions that change over time, we introduce temporal variation in the synergy/discounting parameter, $\omega(t)$.
This reflects another form of changes to resource abundance or wealth as compared to variable rates of return, $r(t)$.

\begin{figure}
\begin{center}
\includegraphics[width=0.95\columnwidth]{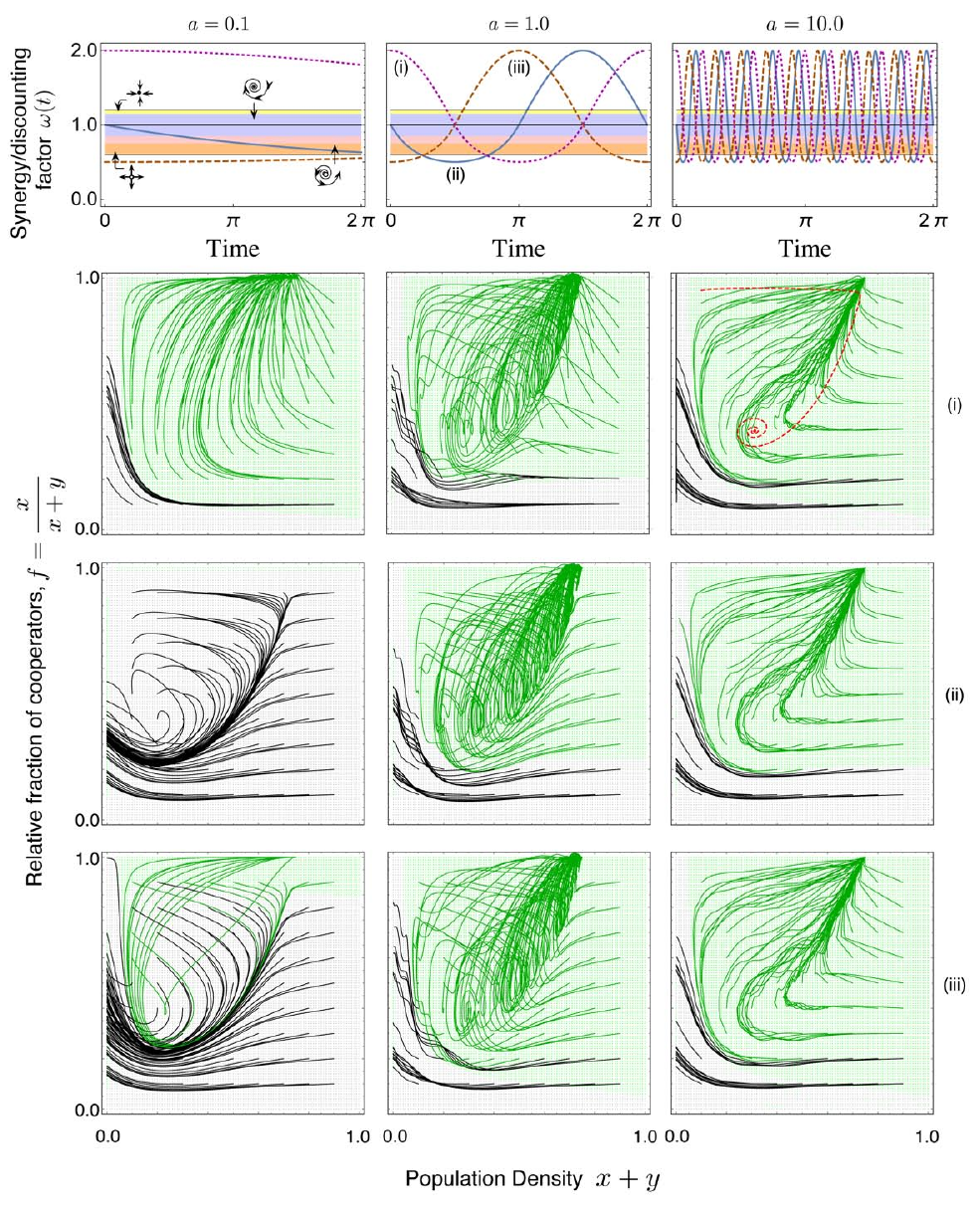}
\caption{
\label{fig:varenv03}
\scriptsize{
\textbf{Eco-evolutionary dynamics under environmental fluctuations: variable synergy/discounting, $\mathbf{\omega(t)}$.}
For periodic oscillations between synergy and discounting, we vary $\omega\in[0.5,2]$ according to \eq{omegat} (top row) and illustrate the dynamics for three timescales $a=\{0.1,1.0,10\}$ (columns), and three phases $\delta=\{\pi/2,0,-\pi/2\}$ (rows (i), (ii) and (iii)). 
For comparison, a sample trajectory (red, dashed) is shown for the dynamics of the mean, $\bar{\omega} = 1$ (panel (i), $a=10$; c.f. \fig{avgdyn}).
Interestingly, in the long run all trajectories either lead to extinction of the entire population (black trajectories) or to the extinction of defectors (green trajectories) resulting in pure cooperator populations, regardless of the initial phase or the timescale of oscillations. Similarly, for sufficiently fast oscillations ($a=1$ and $a=10$) the basins of attraction of each outcome are hardly affected by phase or timescale. 
In contrast, for slow oscillations ($a=0.1$) the basins of attraction sensitively depend on the initial phase and the population likely goes extinct if it takes too long before $\omega$ enters the synergistic regime.
With a negative $a$, if the oscillations are slow i.e. $a=-0.1$, the dynamics in case of (ii) are qualitatively different: as $\omega$ enters the synergistic regime first due to slow oscillations most trajectories have enough time to move towards the all cooperator edge.
Parameters: $N=8$, $r=3$ and $d=0.5$.
}
}
\end{center}
\end{figure}

Implementing periodically oscillating $\omega(t)$ turns out to be more challenging because the range for discounting is bounded, $\omega\in(0,1)$, whereas the range for synergy is not, $\omega\in(1,\infty)$. Because of this asymmetry, we chose
\begin{align}
\label{eq:omegat}
\omega(t) = \begin{cases}
\dfrac1{1+\sin (a t + \delta)}			& \text{if }\sin(a t + \delta)\geq0 \text{ (discounting)}\\ 
1-\sin (a t + \delta)			& \text{if }\sin(a t + \delta)<0 \text{ (synergy)}.
\end{cases}
\end{align}
This ensures that the heaviest discounting, here $1/2$, is counterbalanced by the strongest synergy, here $2$. 
As before, $a$ relates the timescales of environmental fluctuations and eco-evolutionary dynamics and $\delta$ tunes the phase of the oscillation. 
The asymmetry in the range of discounting and of synergy makes the appropriate derivation of the average discounting/synergy, $\bar\omega$, difficult.
Since the arithmetic mean would overestimate the effect of synergy, we choose the geometric mean, which is $\bar\omega \approx 1$.
Interestingly, however, the characteristics of trajectories for $\omega(t)$ turn out to be very different from the dynamics for the average $\bar\omega$ -- regardless of the derivation of $\bar\omega$, see \fig{varenv03} and \fig{avgdyn}. 
For comparable or fast oscillations in $\omega (t)$, most initial conditions lead to homogeneous cooperation at high densities.
This outcome can be attributed to the fact that for most $\omega\in[0.5,2]$ the homogenous cooperator equilibrium is stable. 
However, for slow oscillations the evolutionary outcome becomes highly susceptible to the initial configuration as well as the initial phase of $\omega(t)$, either leading to extinction or homogenous cooperator populations. 
In particular, for very slow oscillations, $a\ll1$, extinction is inevitable because it represents the only stable state for extended periods of time.
This is in stark contrast to the dynamics for the average $\bar\omega = 1$, which suggests persistence of the population and co-existence of cooperators and defectors for most initial configurations \citep{gokhale:WD:2014}.
The discrepancy between the dynamics for the mean $\bar\omega$, see \fig{avgdyn}, and the trajectories for oscillating $\omega$, see \fig{varenv03}, arises because the gradient of selection, $F(f,z)$ is non-linear with respect to $\omega$, which means that the mean of the gradient is not the same as the gradient of the mean, see \app{C} for details. In contrast, the gradient $F(f,z)$ is linear in $r$, which then supports the agreement between the dynamics for $\bar r$ and oscillating rates of return, c.f. \fig{varenv02} and \fig{avgdyn}.

\section{Discussion}
Evolutionary models of social interactions traditionally assume a separation of timescales from ecological processes such that evolutionary selection always acts on ecological equilibria. 
However, ecological `equilibria' may not simply refer to stable population sizes but also oscillatory dynamics based on stable limit cycles and a clear separation of timescales may not always apply. 
Nevertheless, two prominent theoretical frameworks for modelling frequency-dependent evolutionary processes neglect ecological changes: (i) the deterministic replicator dynamics \citep{taylor:MB:1978,hofbauer:book:1998} assumes infinite population sizes and tracks only relative abundances of strategies and (ii) the stochastic dynamics of the (frequency dependent) Moran process \citep{nowak:Nature:2004,moran:book:1962} assumes that the population size is fixed. 
Implicitly this assumes that the carrying capacity is independent of the type and abundance of strategies. 
For a complete understanding of evolutionary processes it is therefore important to incorporate ecological changes. 
Especially evolutionary changes occurring on timescales comparable to ecological changes necessitate an amalgamation into an eco-evolutionary framework \citep{doebeli:PNAS:1997,aviles:EER:1999,aviles:AmNat:2002,mcnamara:Nature:2004,hauert:PRSB:2006,miekisz:LN:2008,bailey:NP:2009}. 
The importance of more comprehensive theoretical approaches is supported by recent experimental results \citep{beaumont:Nature:2009,sanchez:PLoSB:2013}.

We extended the eco-evolutionary framework for linear PG \citep{hauert:PRSB:2006} to include more general forms of social dilemmas by considering non-linear PG interactions through discounted or synergistically enhanced accumulation of benefits \citep{hauert:JTB:2006a}.
We further the study into the ecological domain by considering extrinsic environmental variations, which affect the parameters of social interactions. Intrinsic ecological changes affect the group size in public goods games through variable population densities. This effect is similar to abstaining in public goods interactions, although voluntary participation alone is insufficient to stabilize cooperation and relies on additional mechanisms including spatial structure \citep{hauert:Science:2002}, punishment \citep{hauert:Science:2007} or institutional incentives \citep{sigmund:Nature:2010,sasaki:PNAS:2012}.

Ecological dynamics essentially affects the group size of the public goods game.
As group sizes increase, so do intuitively the possibilities for social conflict. However, the reasons for forming groups may qualitatively change the outcome. For example, defending a resource against a common intruder can reduce social conflict even if group sizes increase\citep{shen:AmNat:2014}. Group size is also essential in foraging \citep{motro:JTB:1991} and variations may promote more egalitarian outcomes in the tragedy of the commune \citep{killingback:BT:2010,brannstrom:JMB:2011}.

Including spatial dimensions either explicitly through unoccupied sites \citep{alizon:Evolution:2008} or implicitly by limiting reproductive opportunities \citep{hauert:PRSB:2006} effectively reduces the interaction group size and shows interesting parallels to voluntary public goods games \citep{hauert:Science:2002}. 
Loners, who do not participate in the public goods game receive benefits between that for mutual cooperation and mutual defection.
An abundance of loners implies smaller interaction group sizes which are favourable for cooperators. As a consequence, the number of participants increases and the public good becomes susceptible to exploitation by defectors. However, once defectors prevail, they are outperformed by defectors and the cycle starts all over again.
Although, the dynamics of such voluntary public goods interactions does not admit stable interior fixed points as opposed to the ecological feedback mechanisms discussed herein -- unless, of course, further mechanisms such as institutionalised incentives come into play \citep{sasaki:PNAS:2012}.

Synergy and discounting generate non-linearities in the rate of return of the PG and hence reflect diminishing returns or economies of scale, which are common features of group interactions in biological and social systems \citep{archetti:JEB:2009,archetti:Evolution:2010,boyd:Science:2007,pena:JTB:2015}. 
For example, in cooperative breeding cichlid fish the optimum breeding group size changes depending on the perceived environmental threats as compared to the potential benefits an additional member could provide to the group \citep{zoettl:PRSB:2013}. 
Additional members can dilute the risk of predation and/or actively take part in territory defence.
Costs due to enhanced brood parasitism, cannibalism and growth reduction \citep{heg:BES:2008,bruintjes:PLoSONE:2011}, however, reduce the benefit leading to active eviction of immigrants \citep{taborsky:Behaviour1985,buston:BE:2003,cant:PRSB:2010}.
In addition, the cichlid example emphasizes an important ecological factor: variable risks. 
In the presence of a predator, being in a group dilutes the risk per individual and also confuses the predator \citep{wrona:AmNat:1991,kokko:PRSB:2001}.
Moreover, it might be possible to actively deter the predator, which would be impossible alone. 
However, larger group sizes also imply larger visibility and higher encounter rates with predators \citep{hebblewhite:CJZ:2002}.
Thus, changes in environmental/ecological factors may alter the characteristics of social interactions and, in turn, affect the evolutionary trajectory, as demonstrated in theory \citep{uyenoyama:TPB:1979,fudenberg:JET:1992,wakano:JTB:2011,libby:PRSB:2011,hanski:ANAS:2012} and experiments \citep{beaumont:Nature:2009,zhang:GB:2010,sanchez:PLoSB:2013}.

Here we considered two sources of ecological variation: intrinsic effects based on population dynamics (\sect{internal}) and extrinsic effects based on changes in the environment (\sect{external}) , which are exemplified by three types of extrinsic, periodic variation in: (i) probabilities to engage either in discounted (diminishing returns) or synergistically enhanced (economies of scale) PG interactions, (ii) efficiency of the PG (varying rate of return, $r$), and (iii) non-linearity in the accumulation of benefits (varying synergy/discounting, $\omega$). 

In the first two cases the characteristics of the trajectories generated under periodical oscillations are in good qualitative agreement with the corresponding average dynamics -- the average of the two games in (i), and the average multiplication factor $\bar r$ in (ii) -- provided that environmental fluctuations are sufficiently fast. 
Interestingly, in scenario (iii) the dynamics based on the average $\bar\omega$ suggests stable co-existence of cooperators and defectors at intermediate population densities, see \citep{gokhale:WD:2014}.
The trajectories under fluctuating $\omega$ converge to high densities of homogeneous cooperator populations. 
More specifically, oscillations in $\omega$ not only promote cooperation but even eliminate defection, provided that the environmental fluctuations arise on a sufficiently fast timescale.
For sufficiently slow oscillations the population will inevitably go extinct in all three cases, if for any value of the oscillating function, extinction is the only stable state. 
The initial configuration and initial phase of environmental oscillations only affects the time and trajectories leading to extinction.

Effects of ecological variation based on intrinsic or extrinsic sources can alter the fitness landscape or, similarly, pleiotropy between traits can change the effective selection pressure observed on a single trait \citep{mcnamara:JRSI:2013}.
Seasonal variation can affect the epidemiology of important vector borne diseases and could have triggered the evolution of plastic transmission strategies that match the temporally varying density of mosquitoes \citep{cornet:PLosPath:2014}.
Even the interaction patterns themselves can be stochastic, furthermore complicating the population dynamics \citep{huang:PNAS:2015}.
Comparisons between data and predictions, which account for such complications, reflects a recent trend in experimental studies \citep{sanchez:PLoSB:2013,lewis:EC:2013}. 
Including non-linear payoffs and temporally fluctuating interaction parameters renders evolutionary game dynamics more complex but provides a window to investigate the rich dynamical scenarios routinely seen in nature. Here we propose ways towards richer evolutionary game theory.

\section*{Acknowledgements}
We thank Christian Hilbe, Bin Wu, and Arne Traulsen for helpful discussions. 
C.S.G. acknowledges financial support from the New Zealand Institute for Advanced Study (NZIAS) and the Max Planck Society. 
C.H. acknowledges the hospitality of the Max Planck Institute for Evolutionary Biology, Pl\"{o}n, Germany and financial support from the Natural Sciences and Engineering Research Council of Canada (NSERC) and the Foundational Questions in Evolutionary Biology Fund (FQEB), grant RFP-12-10.

\renewcommand{\theequation}{A.\arabic{equation}}
\setcounter{equation}{0}

\appendix

\section{Average fitness of cooperators and defectors}
\label{app:A}
\begin{figure}
\begin{center}
\includegraphics[width=0.95\columnwidth]{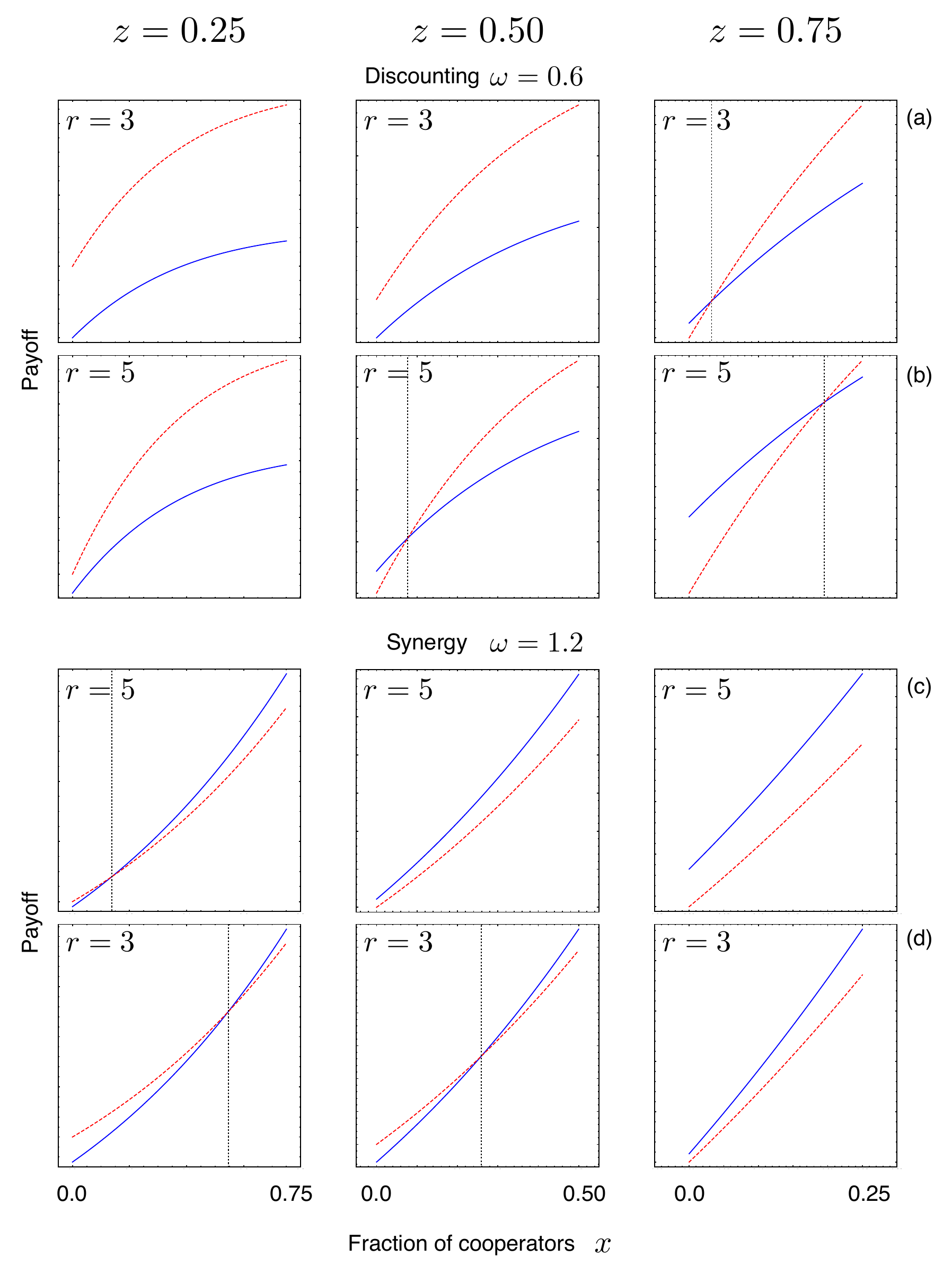}
\caption{
\label{fig:payoffsols}
\scriptsize{
\textbf{Average payoffs for different ecological scenarios.}
The payoff for cooperators, $f_C$ (solid, blue), and defectors, $f_D$ (dotted, red), are shown at high (left, $z=1/4$), middle (centre, $z=1/2$), and low (right, $z=3/4$) population densities, under discounting (rows (a) and (b), $\omega=0.6$) and synergy (rows (c) and (d), $\omega=1.2$), as well as for low (rows (a) and (d), $r=3$) and high (rows (b) and (c), $r=3$) multiplication factors with cost $c=1$ and $N=8$. Together this can generate the four characteristic scenarios of social dilemmas. For example, at $z=1/2$ (centre column), defectors dominate in (a), co-existence in (b), cooperators dominate in (c) (by-product mutualism), and bi-stability in (d) (coordination game).
The characteristics of the game change with population density because low densities support cooperators whereas high densities promote defectors.}}
\end{center}
\end{figure}

The public goods interaction admits up to $N$ players. 
However, at low population densities it may be difficult to recruit $N$ players and hence the effective interaction group size $S$ ranges from $2$ to $N$. 
Note that at least two players are required for social interactions -- a single player gets a zero payoff.
For a focal individual the probability that there are $m$ cooperators among its $S-1$ co-players is given by 
\begin{align}
\binom{S-1}{m} \left(\frac{x}{1-z}\right)^m \left(\frac{y}{1-z}\right)^{S-1-m}.
\end{align}
Setting the costs of cooperation to $c=1$, the payoffs for defectors and cooperators in a group of size $S$ are
\begin{subequations}
\begin{align}
P_D (S) &= \frac{r}{S} \sum_{m=0}^{S-1} \binom{S-1}{m} \left(\frac{x}{1-z}\right)^m \left(\frac{y}{1-z}\right)^{S-1-m} \frac{1-\omega^m}{1-\omega} \\
P_C (S) &= \frac{r}{S}-1 + \frac{r}{S} \sum_{m=0}^{S-1} \binom{S-1}{m} \left(\frac{x}{1-z}\right)^m \left(\frac{y}{1-z}\right)^{S-1-m} \omega\frac{1-\omega^m}{1-\omega}.
\end{align}
\end{subequations}
The probability that an individual interacts in a group of size $S$, i.e. faces $S-1$ co-players, is
\begin{align}
\binom{N-1}{S-1} (1-z)^{S-1} z^{N-S}.
\end{align}
This yields the average payoffs for cooperators and defectors:
\begin{subequations}
\begin{align}
f_D &= \sum_{S=2}^{N} \binom{N-1}{S-1} (1-z)^{S-1} z^{N-S} P_D(S)\\
f_C &= \sum_{S=2}^{N} \binom{N-1}{S-1} (1-z)^{S-1} z^{N-S} P_C(S),
\end{align}
\end{subequations}
which simplify to
\begin{subequations}
\label{eq:fitnesses}
\begin{align}
f_D &= \frac{r}{N}\frac1{1-z-x(1-\omega)} \left[\frac{(x (\omega-1)+1)^{N}-1}{\omega-1} - \frac{x (1-z^N)}{1-z}\right] \\
f_C & = f_D - F(x,z),
\end{align}
\end{subequations}
with
\begin{align}
F(x,z) = 1 + (r-1) z^{N-1} - \frac{r}{N} \frac{(1-x (1-\omega))^{N}-z^N}{1-z-x (1-\omega)}.
\end{align}
In the special case of $\omega = 1$ this reduces to a function in $z$ alone \citep{hauert:PRSB:2006}. 
Effects of fluctuating population densities on the characteristics of evolutionary games can be investigated by considering the fitness of the two strategies, \eq{fitnesses}, at particular densities, see \fig{payoffsols}.
For discounting, $\omega<1$, the resulting interactions are either dominance or co-existence games, whereas for synergy, $\omega>1$, the resulting interactions are either bi-stable or dominance of cooperators (by-product mutualism). 
In either case the population dynamics is capable of triggering a qualitative change in the type of interaction because the population density determines the average interaction group size $S$. At sufficiently low densities $S<r$ holds, which supports cooperators, while at higher densities $S>r$ holds and favours defectors.

\subsection{Homogeneous population}
\label{app:A1}

\begin{figure}[tp]
\begin{center}
\includegraphics[width=\columnwidth]{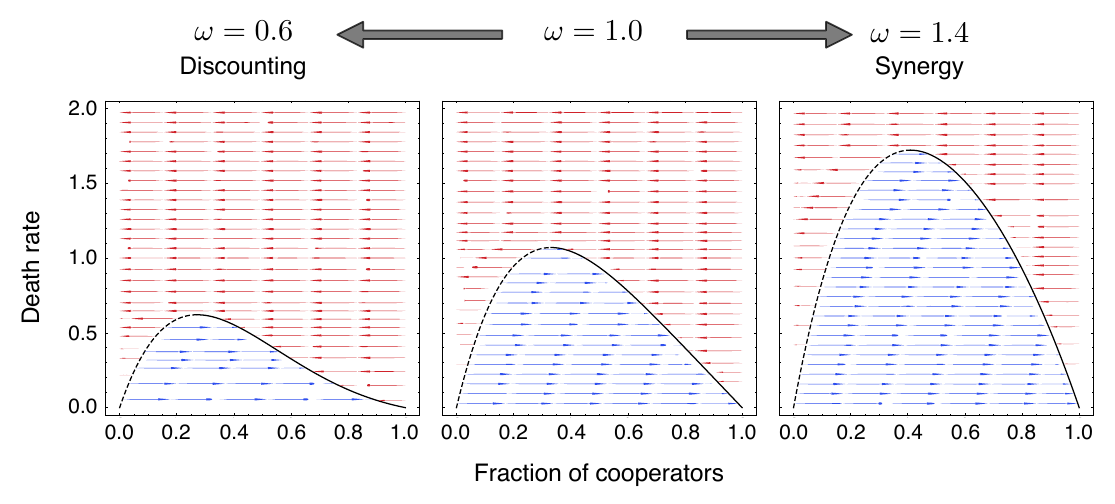}
\caption{
\label{fig:nodefroots}
\scriptsize{
\textbf{Homogeneous cooperator populations.}
In the absence of defectors, $y=0$, a population of cooperators can persist for sufficiently low death rates, $d$.
For $\omega=1$, the maximum is given by $d_\text{max} = (r-1)(N-1) N^{-N/(N-1)}$.
As soon as $d$ exceeds the threshold, $d_\text{max}$, the population goes extinct. The maximum sustainable death rate increases with $\omega$ such that cooperators can sustain higher death rates for synergistically enhanced benefits, $\omega>1$, than for discounted benefits, $\omega<1$.
Solid lines indicate stable population densities and unstable states are indicated by dashed lines.
}}
\end{center}
\end{figure}

If the population consists of only defectors, $x=0$, their average payoff is $f_D = 0$ and we have $\dot{y} < 0$. 
Thus, defectors continue to decrease in abundance and eventually go extinct.
In contrast, in a population of only cooperators, $y=0$, the dynamics of the cooperator density is given by $\dot{x} = x (z f_C -d)$ and their average fitness, $f_C$, from \eq{fitnesses} simplifies to
\begin{align}
f_C = \begin{cases}
(1-r) (1-(1-x)^{N-1}) &\text{if } \omega = 1 \\ 
(1-r) (1-x)^{N-1}+\dfrac{r \left((x(\omega-1)+1)^N-1\right)}{N (\omega -1)x}-1 & \text{otherwise} .
\end{cases}
\end{align}
Apart from the trivial fixed point $x=0$, which marks extinction, further fixed points may exist whenever $d = z f_C$, see \fig{nodefroots}.
For $\omega = 1$ an explicit expression for the maximum death rate that a population of only cooperators can sustain is given by $d_\text{max} = (r-1) (N-1) N^{-N/(N-1)}$. 
For $d>d_\text{max}$ the population invariably goes extinct. For $d<d_\text{max}$ the population may persist provided that the initial population density is sufficiently high.
Unfortunately, for general $\omega$ (and $N$) analytical expressions for $d_\text{max}$ are not accessible. 
However, for $\omega>1$ cooperators can sustain greater death rates while the converse holds for $\omega<1$, see \fig{nodefroots}.
If the density of cooperators is low then the benefits produced by the public good are unable to offset the death rate and the population goes extinct. 
The threshold density required to sustain the population decreases with increasing $\omega$, i.e. moving from discounted to synergistically enhanced benefits. 
In ecology, density dependent effects reflecting difficulties in finding interaction partners (or mates) are referred to as the Allee effect \citep{stephens:TREE:1999}.

\section{Fixed points and stability}
\label{app:B}
\begin{figure}[tp]
\begin{center}
\includegraphics[width=\columnwidth]{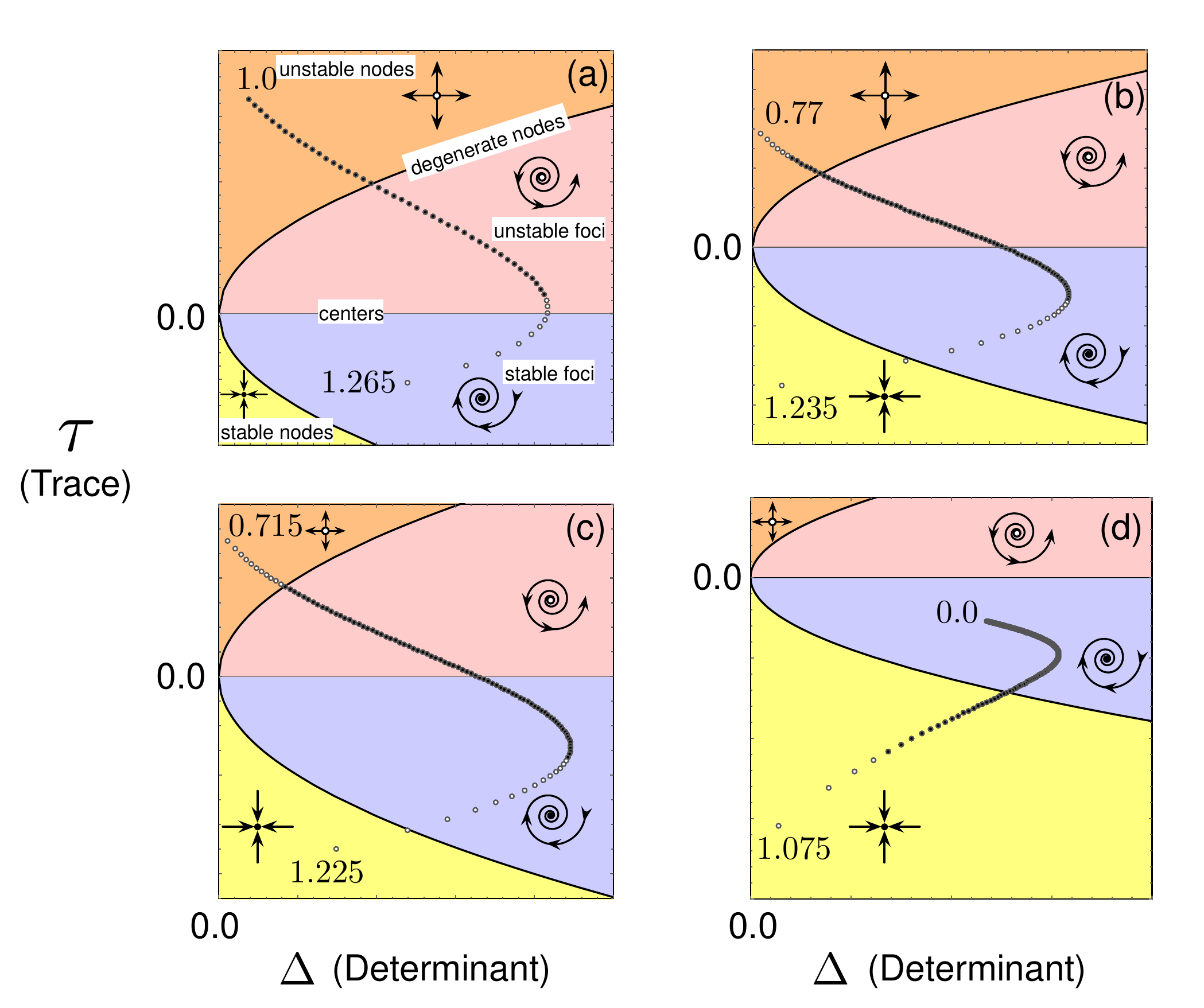}
\caption{
\label{fig:phasep}
\scriptsize{
\textbf{Stability analysis of the interior fixed points while varying the synergy/discounting parameter $\omega$.}
Parameters are the same as in \fig{allplots}: $N = 8$, $d=0.5$
(a) $r = 2.2$
(b) $r = 2.7$
(c) $r = 2.8$
(d) $r = 5$.
For $\omega = 1$ at most a single fixed point $\bf{Q}$ can exist in the interior of the $(1-z,f)$-phase plane.
Varying the synergy/discounting parameter $\omega$, another fixed point $\bf{P}$ may appear. 
Whenever $\bf{P}$ exists, it is always a saddle, $\Delta<0$. Therefore we track only the stability and dynamical characteristics of $\bf{Q}$ by following its trajectory through the space spanned by the determinant $\Delta$ and the trace $\tau$ of the Jacobian matrix, \eq{J}, as a function of $\omega$. The curves $\tau=0$ and $\tau^2 -4 \Delta =0 $ (parabola) divide the space in four regions with different dynamical features as indicated in panel (a). Typically, $\bf{Q}$ appears as an unstable node at low $\omega$. As $\omega$ increases, $\bf{Q}$ first turns into an unstable focus, then a stable focus and usually ends up as a stable node before disappearing again. Note that not all dynamical regimes may be observed for all parameters. For example, in (d) $\bf Q$ already appears as a stable focus. The presence of the second fixed point $\bf{P}$ is indicated by open circles.
}
}
\end{center}
\end{figure}

In order to determine the stability of the interior fixed points, $\bf Q$ and $\bf P$, we need to resort to numerical evaluations of the trace, $\tau$, and determinant, $\Delta$, of the Jacobian of \eq{alteredfz} at each interior fixed point. The interior fixed points are given by non-trivial solutions to $\dot f=0$ and $\dot z=0$, see \eq{alteredfz}. From $\dot z=0$ follows that 
\begin{align}
\label{eq:fateq}
f_\text{eq} = \frac{d}{(r-1) z (1-z^{N-1})}
\end{align}
and similarly, $\dot f=0$ requires that $F(f_\text{eq},z)=0$ where
\begin{align}
\label{eq:fzf}
F(f_\text{eq}, z) =&\ 1 + (r-1) z^{N-1} - \frac{r}{N} \frac{(1-f_\text{eq} (1-z) (1-\omega))^{N}-z^N}{1-z-f_\text{eq} (1-z) (1-\omega)},
\end{align}
which implicitly defines $z_\text{eq}$ and may admit several solutions in $[0,1]$. Unfortunately $z_\text{eq}$ is analytically inaccessible. Numerical analysis shows that depending on $r$ and $N$ there are zero, one or two solutions, which corresponds to no interior fixed point, one fixed point $\bf Q$ or two fixed points $\bf Q$ and $\bf P$. Calculating the Jacobian at $\bf Q$ and $\bf P$ using $F(f_\text{eq},z_\text{eq})=0$ and \eq{fateq} yields,
\begin{align}
\label{eq:J}
\mathbf{J} =
\begin{pmatrix}
-z f (1-f) \frac{\partial F(f,z)}{\partial f} & 
-z f (1-f) \frac{\partial F(f,z)}{\partial z} \\
-(r-1) (1-z) z (1-z^{N-1}) & 
-(r-1) (1-z) f (1-N z^{N-1})
\end{pmatrix}.
\end{align}
The trace $\tau$ and determinant $\Delta$ are then given by,
\begin{align}
\tau =&\ -f\left[ (r-1)(1-z)\left(1-N z^{N-1}\right) + z (1-f) \frac{\partial F(f,z)}{\partial f}\right], \\
\Delta =&\ f(1-f) z(1-z) (r-1) \left[ f\left(1-N z^{N-1}\right) \frac{\partial F(f,z)}{\partial f} - z\left(1-z^{N-1}\right) \frac{\partial F(f,z)}{\partial z}\right].
\end{align}
Numerical evaluations of $\tau$ and $\Delta$ for both internal fixed points $\bf{P}$ and $\bf{Q}$ reveal that $\bf{Q}$ can exhibit a variety of dynamical properties but interestingly, $\bf P$, whenever present, is always a saddle point, $\Delta < 0$. 
As $\omega$ changes, the fixed point $\bf{Q}$ follows a curve through the space spanned by $\tau$ and $\Delta$, see \fig{phasep}. 

For strong discounting (small $\omega$) no interior fixed point exists and the population goes invariably extinct. 
As $\omega$ increases one interior fixed point may appear through a transcritical bifurcation or two fixed points through a saddle node bifurcation. 
The presence, location and stability of the fixed points $\bf Q$ and $\bf P$ is analytically inaccessible (except when $\omega = 1$ \citep{hauert:PRSB:2006,hauert:TPB:2008}) and has been determined numerically \fig{phasep}. 
Since in the absence of synergy or discounting, $\omega=1$, at most a single interior fixed point exists, the second interior fixed point disappears through a transcritical bifurcation (leaving the $(1-z,f)$-plane), while $\omega$ is still in the discounting regime. The interior fixed point then undergoes a Hopf-bifurcation -- either sub- or super-critical depending on the game parameters \citep{hauert:TPB:2008}. 
In the synergistic regime, a second interior fixed may (again) appear and for still higher $\omega$ the two interior fixed points collide and disappear in another saddle node bifurcation or one interior fixed point disappears through a transcritical bifurcation. 
Finally, for strong synergistic effects defectors always go extinct leaving a homogenous population of cooperators behind. 
Interactive simulations provide opportunities for further online explorations of the rich eco-evolutionary dynamics \citep{gokhale:WD:2014}.

Typically, $\bf Q$ passes through various phases of stability (illustrated by the cartoons for local stability) -- starting as an unstable node at low $\omega$, turning into an unstable focus, then a stable focus and finally into a stable node before $\bf Q$ disappears at high $\omega$. Since $\tau$ is polynomial in $\omega$ and $\omega$ is continuous, the fixed point becomes a centre (degenerate focus) for particular values of $\omega$. 

\section{Average dynamics under oscillating environments}
\label{app:C}

\begin{figure}[tp]
\begin{center}
\includegraphics[width=\columnwidth]{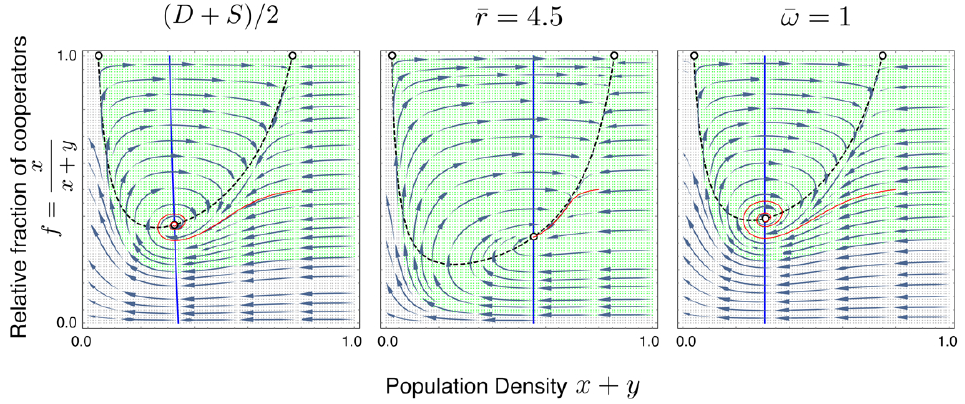}
\caption{
\label{fig:avgdyn}
\scriptsize{
\textbf{Dynamics under average of the environmental oscillations}
Taking the average of the environmental oscillations allows us to analytically evaluate the dynamics and the fixed points.
For a variation in interaction types, the discounting and synergistic scenarios oscillate with probability $p_D(t)=(\sin(a t) + 1)/2$. Here taking the average value of $p_D(t)=0.5$ such that we have $(D+S)/2$ the dynamics is visualised in the left panel with relevant parameters being $\omega_D=0.9$, $\omega_S=1.1$, $r_D=4.2$, $r_S=2.1$.
For a variation in the rate of return, oscillating as per $r(t)=3 \sin (a t) + 4.5$, the average is $\bar{r}=4.5$ as shown in the central panel with $\omega =1$.
For a variation in the synergy/discounting parameter, oscillating as per \eq{omegat}, the geometric average is $\bar{\omega}=1$ resulting in the dynamics as visualised in the right panel for $r=3$.
For all scenarios, we have $N=8$ and $d=0.5$.
}
}
\end{center}
\end{figure}

The trajectories calculated using the oscillating functions have been shown in the main text for (i) variation in the interaction types, \fig{varenv01}; (ii) variation in the rate of return, \fig{varenv02}; and (iii) variation in the synergy/discounting factor, \fig{varenv03}.
Typically assuming a separation of timescales between the faster environmental oscillations and the slower evolutionary dynamics, the environmental effects can be averaged out.
Here we show the dynamics, which emerges following this assumption in \fig{avgdyn}.

\begin{figure}[tp]
\begin{center}
\includegraphics[width=\columnwidth]{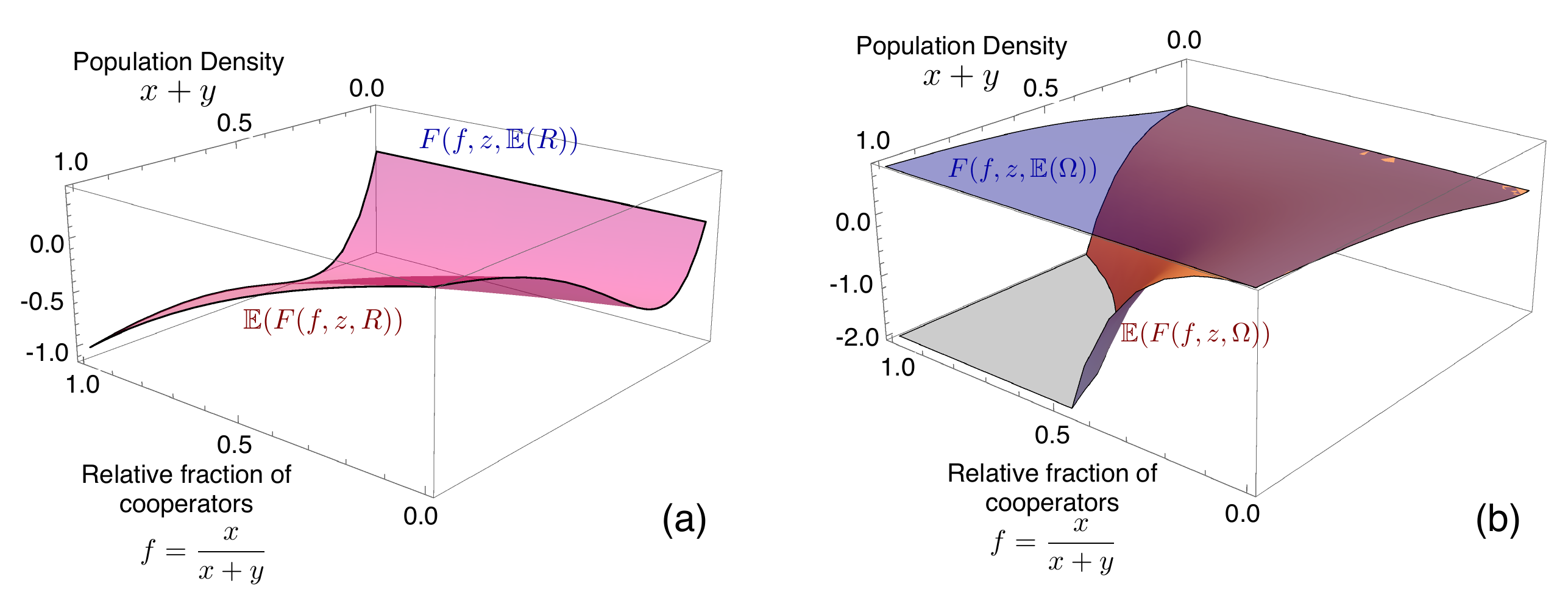}
\caption{
\label{fig:jensen}
\scriptsize{
\textbf{Illustration of Jensen's inequality}
(a) For Gaussian distributed rates of return, $r$, with mean $\bar r=4.5$ and variance $1$ the payoff difference $F(f,z) = f_D - f_C$ of the mean equals the mean of the payoff difference, $\mathbb{E} (F(f,z,R)) = F(f,z,\mathbb{E}(R))$ because it linearly depends on $r$.
(b) In contrast, for Gaussian distributed $\omega$ the function of the mean $F(f,z)[\mathbb{E}(\Omega)]$ (blue, translucent surface) differs from the mean of the function $\mathbb{E} (F(f,z)[\Omega])$ (red, solid surface)}. The latter turns out to be consistently smaller and hence fluctuations in $\omega$ favour cooperators.
Parameters: $N=8$ (a) $\omega=1.2$ (b) $r=2$.
}
\end{center}
\end{figure}

For fast and even comparable timescales between the oscillations in the interaction types and the rate of return we do indeed see the trajectories reflecting qualitatively similar dynamics as that for the average (compare \figs{varenv01}{varenv02} with left and middle panels in \fig{avgdyn}).
For oscillating synergy/discounting $\omega$, however, neither fast nor comparable timescales recover the dynamics for the average, $\bar\omega$ (compare \fig{varenv03} and \fig{avgdyn}, right panel). 
One reason that the dynamics is well captured by the average in the case of oscillating rates of returns, $r$, but not for oscillating synergy/discounting, $\omega$, is that the gradient of selection $F(f,z)$ is a linear function of $r$ whereas it is nonlinear in $\omega$. In order to illustrate this difference, we consider Jensen's inequality, which states that the average of a non-linear function is different from the function evaluated at the average of a random variable \citep{jensen:AM:1906,gillespie:AmNat:1977}. More specifically, we consider two Gaussian random variables, $R$ and $\Omega$, centered around $\bar r=4.5$ and $\bar\omega=1$, respectively, with variance $1$. To avoid meaningless negative values, the distribution is symmetrically truncated at $0$. From the linearity in $r$ follows that $\mathbb{E}(F(f,z)[R]) = F(f,z)[\mathbb{E}(R)]$ for fixed $N$ and $\omega$. As a consequence the dynamics for fluctuating $r$ matches that of $\bar r$, provided that fluctuations arise on sufficiently fast time scales. In contrast, $\mathbb{E}(F(f,z)[\Omega]) \neq F(f,z)[\mathbb{E}(\Omega)]$ for fixed $N$ and $r$, see \fig{jensen}. It turns out that the function of the mean exceeds the mean of the function, $\mathbb{E}(F(f,z)[\Omega])<F(f,z)[\mathbb{E}(\Omega)]$. Since $F(f,z)$ denotes the advantage of defectors over cooperators, it follows that fluctuations in $\omega$ are beneficial for cooperation and has been verified for various $r$.

If environmental variations occur at a slower timescale than the evolutionary dynamics then the results are drastically different from the averages (compare left columns in Figures \figs{varenv01}{varenv02} and \ref{fig:varenv03} with the averages in \fig{avgdyn}). 
In these cases even the phase, in which the system enters the particular scenarios, is important for the trajectories eventual unravelling.


%

\end{document}